\documentclass{osa-article}

\journal{oe}

\begin{document}

\title{Rotation measurements using a a resonant fiber optic gyroscope based on Kagome fiber}

\author{Alexia Ravaille,\authormark{1,2,3} Gilles Feugnet,\authormark{2} Beno\^it Debord,\authormark{4} Fr\'ed\'eric G\'er\^ome,\authormark{4} Fetah Benabid,\authormark{4} and Fabien Bretenaker\authormark{1,5,*}}

\address{\authormark{1}Laboratoire Aim\'e Cotton, ENS Paris-Saclay, Universit\'e Paris-Sud, CNRS, Universit\'e Paris-Saclay, Orsay, France\\
\authormark{2}Thales Research \& Technology, Palaiseau, France\\
\authormark{3} Thales Avionics, Ch\^atellerault, France\\
\authormark{4} GPPMM Group, XLIM Research Institute, CNRS, Universit\'e de Limoges, Limoges 87032, France\\
\authormark{5} Light and Matter Physics Group, Raman Research Institute, Bangalore 560080, India}

\email{\authormark{*}fabien.bretenaker@u-psud.fr} 



\begin{abstract}
We build a resonant fiber optic gyro based on Kagome hollow-core fiber. A semi-bulk cavity architecture based on a 18-m-long Kagome fiber permits to achieve a cavity finesse of 23 with a resonance linewidth of 700\;kHz. An optimized Pound-Drever-Hall servo-locking scheme is used to probe the cavity in reflection. Closed-loop operation of the gyroscope permits to reach an angular random walk as small as 0.004$^\circ/\sqrt{\mathrm{h}}$ and a bias stability of 0.45$^\circ$/h over 0.5~s of integration time. 
\end{abstract}

\section{Introduction}
Thanks to its resonant cavity, the resonant fiber optic gyroscope (RFOG) has the potential to achieve high performance rotation sensing (navigation grade) with up to 100 times shorter fiber length than the interferometric fiber optic gyroscope (IFOG) \cite{MaHuilian2010}. This size advantage makes the RFOG a serious alternative to the IFOG. However, the RFOG has yet to achieve a level of performance comparable with IFOGs, despite the large progress made and efforts undertaken in its development. A recent example of state-of-the art result is the work of Sanders et al. who have built an RFOG based on standard fiber, reaching an angular random walk (ARW) of 0.0029 $^\circ/\sqrt{\mathrm{h}}$ and a bias stability of 0.02$^\circ$/h \cite{Sanders2017(2)}. Among the chief challenges that are left to accomplish for the RFOG to reach navigation performance-grade, we count the reduction of the bias drift induced by Kerr effect and temperature driven polarization instabilities  \cite{Iwatsuki1986,Iwatsuki1986(2)}.

This explains why the recent progress in hollow-core photonic crystal fibers has renewed the interest in RFOG development \cite{Sanders2006}. Indeed, it has been shown that using HC-PCF to build the sensing coil in an RFOG permits light to propagate mainly in air, thus making the RFOG less sensitive to such Kerr non-linearities and temperature effects \cite{Sanders2006,Terrel2012}.

Conversely, the use of HC-PCF in a resonant cavity has its own drawbacks. The main one is that there isn't yet a HC-PCF coupler allowing to build a low loss resonant cavity out of the HC-PCF alone. There are currently two main ways to implement a HC-PCF in a RFOG. The first one is to directly splice the HC-PCF to a PM fiber coupler \cite{Terrel2012,YanYuchao2015,YanYuchao2015(3)}. However, the splice-losses ($\sim 2$ dB per splice) reduce the finesse of the resonant cavity below 10. In this case, the best performances achieved to date exhibit an angular random walk (ARW) equal to 0.055$^\circ/\sqrt{\mathrm{s}} = 3.3^\circ/\sqrt{\mathrm{h}}$ \cite{Terrel2012} and a bias drift of 0.07$^\circ$/s = 250$^\circ$/h over 4 400~s integration time \cite{YanYuchao2015}. Such noises  are 10,000 times larger than the requirements for navigation grade performances \cite{Lefevre2014}
. The strong losses inside the cavity, which degrade the finesse,  and the reflection at the fiber splices, seem to be the main causes of those relatively low performances.

The other way to implement a HC-PCF inside a resonant cavity is to use free space optics \cite{Sanders2006,JiaoHongchen2017,JiaoHongchen2017(2)} to close the resonant cavity. In this case, the resonator finesse can be as large as 40 \cite{Sanders2006}. The best performances achieved to date with this technology correspond to an ARW of 0.075$^\circ/\sqrt{\mathrm{h}}$ \cite{JiaoHongchen2017} and a bias stability of 2$^\circ$/h over 300~s of integration time \cite{JiaoHongchen2017(2)}, with a finesse equal to 12 only. According to the authors of this latter work, the polarization noise, which is partially suppressed by an adapted modulation, is the main source of bias drift in their gyroscope. One can also wonder whether the servo-loop dynamic is also a source of performance degradation because the cavity is probed in transmission and not in reflection. As a result, Pound-Drever-Hall (PDH) servo-loop control scheme \cite{Black2001}, which has been proven to give better results in terms of RFOG bias stability \cite{JinXiaojun2018}, can be utilized.

There are two  technologies available to build low-losses HC-PCFs, namely the fibers based on photonic bandgap (PBG) structures \cite{Cregan1999} and the so-called inhibited coupling (IC) HC-PCFs \cite{Couny2007(2)}. The latter ones have the advantage to exhibit larger mode diameters, making low-loss coupling to free space propagating beams easier, and to be immune to contamination. Besides, we have recently shown that IC Kagome fibers can be used to build resonators with a good finesse and exhibiting a resonance contrast close to unity \cite{Fsaifes2016}. The aim of the present paper is thus to demonstrate a new HC-PCF RFOG based on such a Kagome fiber. We choose to use free space optics to close the resonant loop to avoid splice losses and backscattering. 
We also choose to implement a PDH servo-loop technique which allows high frequency modulation and hence high closed-loop bandwidth for the frequency servo-locking. 

\begin{figure}[h!]
\centering\includegraphics[width=0.8\textwidth]{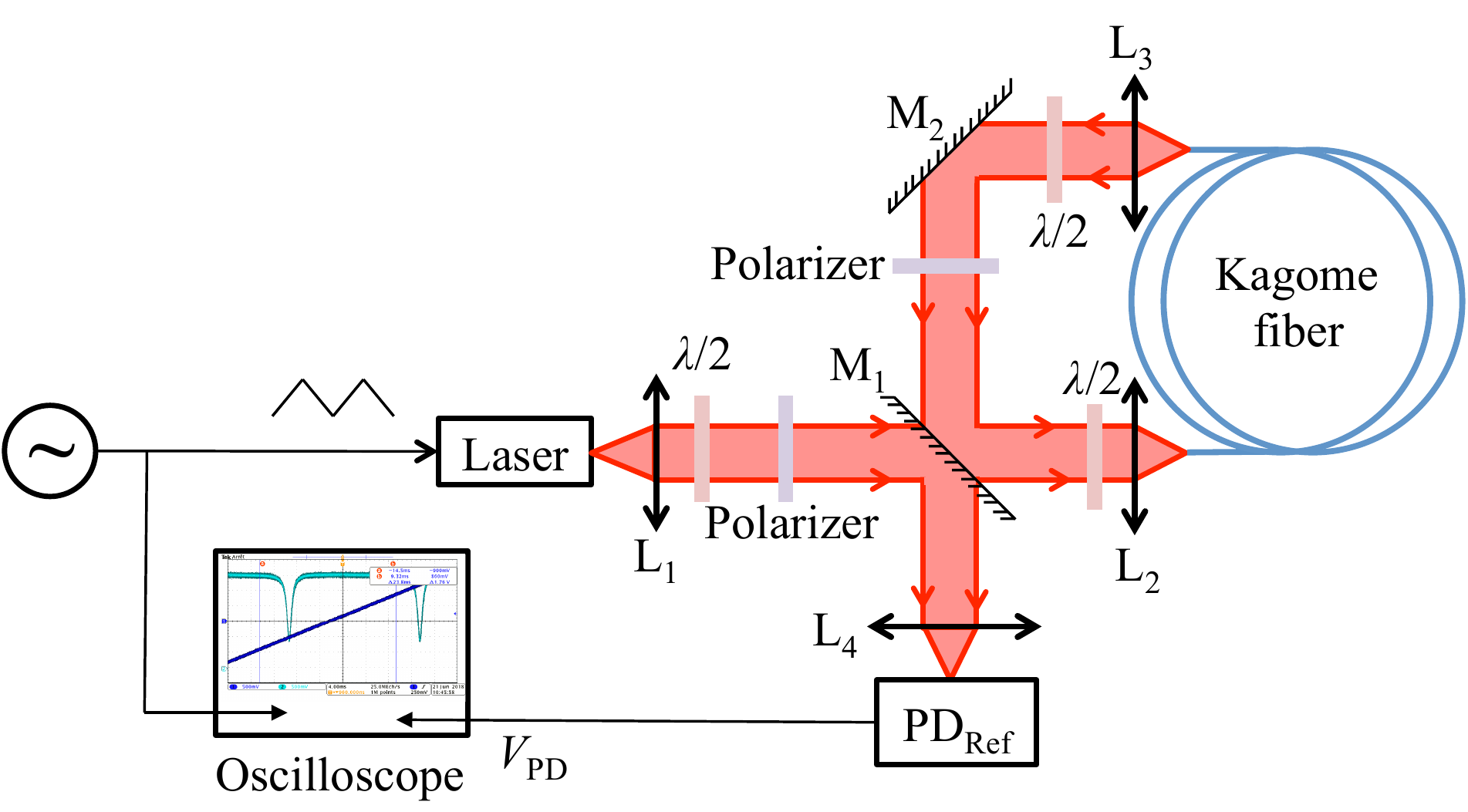}
\caption{\label{Fig_MontageCavResCarac} Experimental setup used to measure the characteristics of the fiber resonator. The laser frequency can be linearly swept to probe the resonator resonances. PD: photodetector.}
\end{figure}
\section{Resonant cavity characterization}
\label{Sec_CaracCav}
The fundamental limitation of the ARW of RFOGs comes from the shot noise in the photocurrent generated by the photodiode that detects the light \cite{Ezekiel1978}. As shown in \cite{Feugnet2017} in the case of a Pound-Drever-Hall (PDH) servo-loop with optimized phase modulation amplitude of $\beta$=1.08~rad, critical coupling, and low-loss cavity, the shot noise limited ARW $\delta\dot{\theta}_\mathrm{SNL}\sqrt{\tau}$ is given by :
\begin{equation}
\label{Eq_SNL}
    \delta\dot{\theta}_\mathrm{SNL}\sqrt{\tau} = \dfrac{\Delta\nu_\mathrm{FSR}}{\mathcal{F}D}\sqrt{\dfrac{\lambda hc}{2\eta P_0}}\ ,
\end{equation}
where $D$ is the diameter of the fiber sensing coil, $\Delta\nu_\mathrm{FSR}$ is the free spectral range of the cavity, $\mathcal{F}$ its finesse, $\lambda$ the wavelength of the laser, $\eta$ the quantum efficiency of the detector, $\tau$ the integration time of the detection and $P_0$ the optical power incident  on the cavity.
As can be seen from eq.\;(\ref{Eq_SNL}), the higher the finesse, the better the sensitivity. We choose to build a semi-bulk resonant optical cavity, as shown in Fig.\;\ref{Fig_MontageCavResCarac}, in order to reach a good enough finesse and to allow an easy change of the hollow-core fiber if needed.

The hollow-core fiber we use is a 18-m-long Kagome fiber . It has a core diameter of approximately 60 $\mu$m and an attenuation of 12~dB/km at 1.55~$\mu$m. 

In order to minimize the fiber bend loss, the fiber is wound with a 30 cm radius coil. The construction of the cavity is completed by plane mirrors M$_1$ and M$_2$, which are placed in a manner to recirculate the output light from each of the fiber ends back into the fiber. Mirror M$_2$ is highly reflecting and mirror M$_1$ has a transmission of 5.8\;\% and a reflection coefficient of 93.9\;\% for the s polarization.  The two intra-cavity lenses L$_2$ and L$_3$ are anti-reflection coated at 1.5\;$\mu$m and have a focal length equal to 40\;mm. In order to mitigate the depolarization induced by the fiber propagation, we use polarizers and half-wave plates to control the intra-cavity polarization. This permits to align the polarization of the resonant mode with the s polarization on the two mirrors. This minimizes  the losses of this resonant mode and permits to excite only one polarization mode of the cavity by injecting s-polarized light. The laser we use is a tunable Koheras laser from NKT Photonics. Lenses L$_1$ and L$_4$ used to match the laser beam to the cavity mode and to focus the reflected beam to the detector, respectively, have a focal length of 11\;mm.

\begin{figure}[h!]
\centering\includegraphics[width=0.8\textwidth]{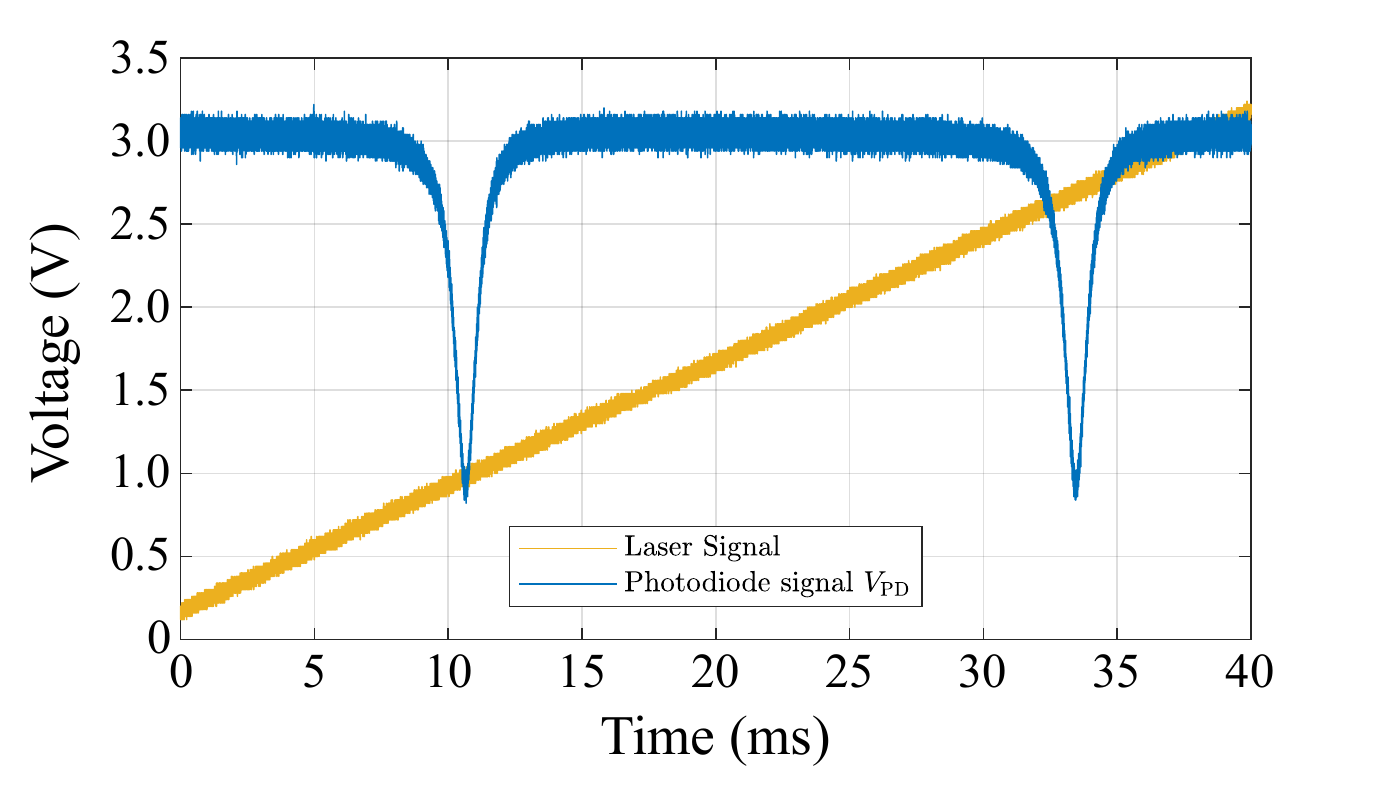}
\caption{\label{Fig_FitCav} Experimental evolution of the power reflected by the cavity when the incident light frequency is linearly scanned.}
\end{figure}

The reflection spectrum we obtain from this cavity is reproduced in Fig.\;\ref{Fig_FitCav}. One can see in particularly that it contains only one series of peaks, showing that all spurious modes have been efficiently eliminated. This spectrum leads to a measurement of a cavity  finesse equal to $\mathcal{F} = 23$, of a free spectral range equal to $\Delta\nu_\mathrm{FSR} = 16.35$~MHz, corresponding to a cavity linewidth $\Gamma \simeq 700$~kHz.

From these numbers, assuming a $P_0 = 1$~mW power incident on the cavity, the shot noise limit is predicted from eq. (\ref{Eq_SNL}) to correspond to an angular random walk equal to $\mathrm{ARW} = \delta\dot{\theta}_\mathrm{SNL}\sqrt{\tau} = 6 \times 10^{-5} ~^\circ/\sqrt{\mathrm{h}}$. From Fig.\;\ref{Fig_FitCav}, one can see that the cavity is not exactly critically coupled, because the reflected intensity at resonance is not exactly zero. Thus, expression (\ref{Eq_SNL}) is not strictly valid, but provides a good approximation of the maximum performances attainable by such a gyroscope.

\section{Pound-Drever-Hall servo-loop optimization}

In order to be able to measure the rotation rate with the RFOG, one has to servo-lock the cavity to the laser, or vice versa, for at least one propagation direction.
There are currently two main ways to implement such a servo-loop. The first one is to use the frequency modulation/demodulation technique \cite{Bjorklund1983}, that consists in modulating the phase or the frequency of the laser at a frequency lower than the cavity linewidth $\Gamma$. After demodulation of the detected power, this technique allows to monitor either the transmission or the reflection of the cavity. But the bandwidth of the servo loop is hence limited by this modulation frequency, which is of the order of a few kHz.
The second way is to use the PDH technique \cite{Black2001}, which consists in modulating the phase of the laser at a frequency much larger than the cavity linewidth. This technique can be implemented only by detecting the power reflected by the cavity because the sidebands produced by the phase modulation are out of resonance and are thus fully reflected by the cavity. This technique allows to reach a much broader servo-loop bandwidth than the preceding one, because the modulation frequency is usually in the MHz range.

\begin{figure}[h!]
\centering\includegraphics[width=0.8\textwidth]{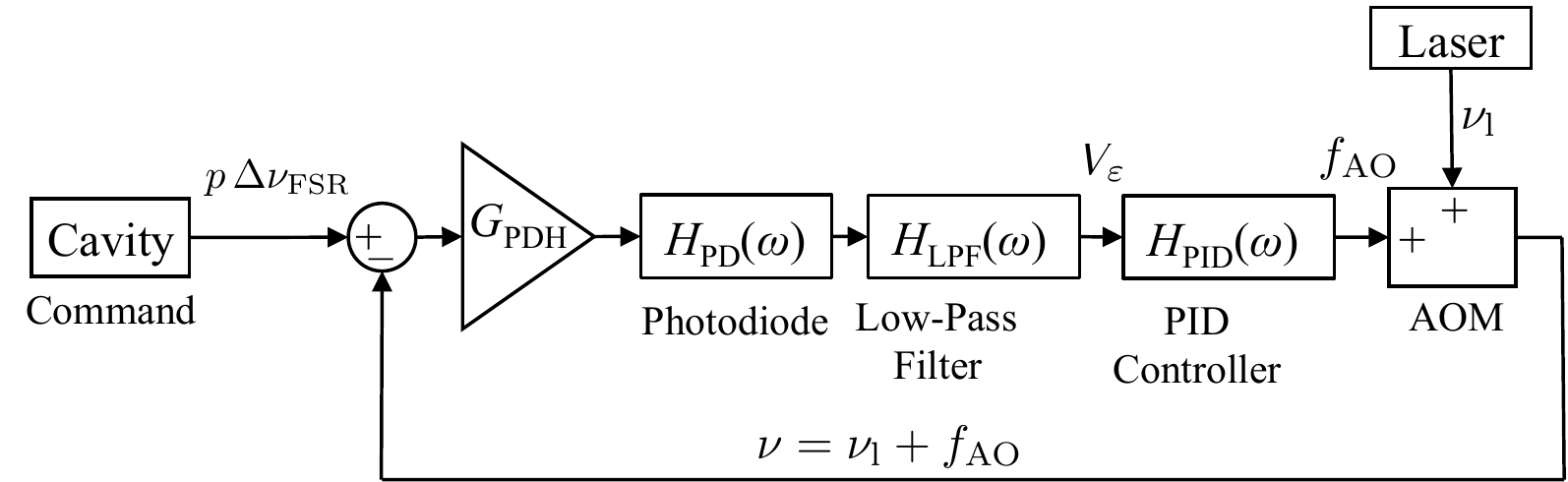}
\caption{\label{Fig_MOD_PDH}Model of the PDH servo-loop control. The resonance frequency of the cavity is modeled by a command of an integer multiple $p$ of the free spectral range $\Delta\nu_{\mathrm{FSR}}$.}
\end{figure}
As will be detailed in Section \ref{Rotation}, we implemented two PDH servo-loops to track the rotation rate of our device. These PDH servo-loops are based on two acousto-optic modulators (AOMs) that keep the light injected in the two counterpropagating directions inside the cavity at resonance with the chosen cavity modes. In order to optimize these PDH servo-loops, we designed the model schematically shown in Fig. \ref{Fig_MOD_PDH}. It is based on a phase modulator (PM) driven at frequency $f_m$, the resonant cavity probed in reflection, a photodiode (PD) to detect the intensity reflected by the cavity, a demodulator that demodulates the photodiode signal at $f_m$, a low-pass filter  producing the error signal ($V_\varepsilon$) and a controller (here a digital PID controller) that applies feedback on the AOM to adjust the injected laser frequency. 

As derived in \cite{Black2001}, when the laser frequency is close to a resonance frequency of the cavity, the error signal $V_\varepsilon$ evolves linearly with the detuning between the incident optical frequency  $\nu_\mathrm{l}+f_\mathrm{AO}$ and the cavity resonance frequency $p\,\Delta\nu_\mathrm{FSR}$:
\begin{equation}
    V_\varepsilon = G_\mathrm{PDH}\ \delta\nu\ ,
\end{equation}
where $\delta\nu = \nu_\mathrm{l}+f_\mathrm{AO}-p\,\Delta\nu_\mathrm{FSR}$ and $G_\mathrm{PDH}$ is the DC gain of the detection chain that is related with the detected laser power, the photodiode, and the demodulator. This is why the sub-system \{PM + PD + demodulator + low-pass filter\} is modelled by a simple frequency discriminator, schematized as a subtractor in Fig. \ref{Fig_MOD_PDH}. The gain factor $G_\mathrm{PDH}$ can be experimentally measured by monitoring the error signal when the controller is off and when the laser frequency is swept.

The photodetectors we use (Thorlabs PDA-10CS-EC) are modelled by a first order low-pass filter with a cut-off frequency $f_{\mathrm{c,PD}}$ :
\begin{equation}
    H_\mathrm{PD}(\omega)=\dfrac{1}{1+\dfrac{\mathrm{i}\omega}{2\pi f_{\mathrm{c,PD}}}}\ ,
\end{equation}
where $\omega$ is the angular frequency of the considered signal. The demodulation and filtering are performed by a digital lock-in amplifier (Z\"{u}rich Instrument UHF-LI).
The demodulation filter is a second-order low-pass filter a with cut-off frequency $f_{\mathrm{c,LPF}}$. It is numerically implemented by the UHF-LI unit, with an equivalent analog transfer function analog domain given by:
\begin{equation}
     H_\mathrm{LPF}(\omega)=\dfrac{1}{1+\dfrac{i\omega\sqrt{\sqrt{2}-1}}{2\pi f_{\mathrm{c,LPF}}}}\ .
\end{equation}

The PID controller is also implemented by the UHF-LI and has a transfer function given by:
\begin{equation}
H_\mathrm{PID}(\omega) = G_\mathrm{P} + \dfrac{G_\mathrm{I}}{i\omega} + \dfrac{G_\mathrm{D}G_\mathrm{N}}{1+\dfrac{G_\mathrm{N}}{i\omega}},
\end{equation}
where $G_\mathrm{P}$, $G_\mathrm{I}$, and $G_\mathrm{D}$ are the proportional, integral, and derivative gain coefficients, respectively, and $G_\mathrm{N}$ is the clamping gain that prevents the derivative controller from diverging at high frequencies. The overall open loop transfer function $H_\mathrm{OL}(\omega)$ of the servo-loop is given by :
\begin{equation}
H_\mathrm{OL}(\omega) = G_\mathrm{PDH}\, H_\mathrm{PD}(\omega)\, H_\mathrm{LPF}(\omega) \, H_\mathrm{PID}(\omega),
\end{equation}
and the closed loop transfer function $H_\mathrm{CL}(\omega)$ reads:
\begin{equation}
H_\mathrm{CL}(\omega) = \dfrac{H_\mathrm{OL}(\omega)}{1+H_\mathrm{OL}(\omega)}.
\end{equation}

Contrary to what has been done in \cite{MaHuilian2012(2)}, we neglect in our simulation the response delay of the loop, which we measured to be equal to $2\;\mu$s. The main difficulty in implementing a PDH servo-loop is to properly choose the controller gains in order to obtain the fastest yet not oscillating response. In the following, we simulate the step response of our system and try to adjust the PID gain coefficients, knowing all the other parameters that enter in the modelling.

\begin{figure}[h!]
\centering\includegraphics[width=0.8\textwidth]{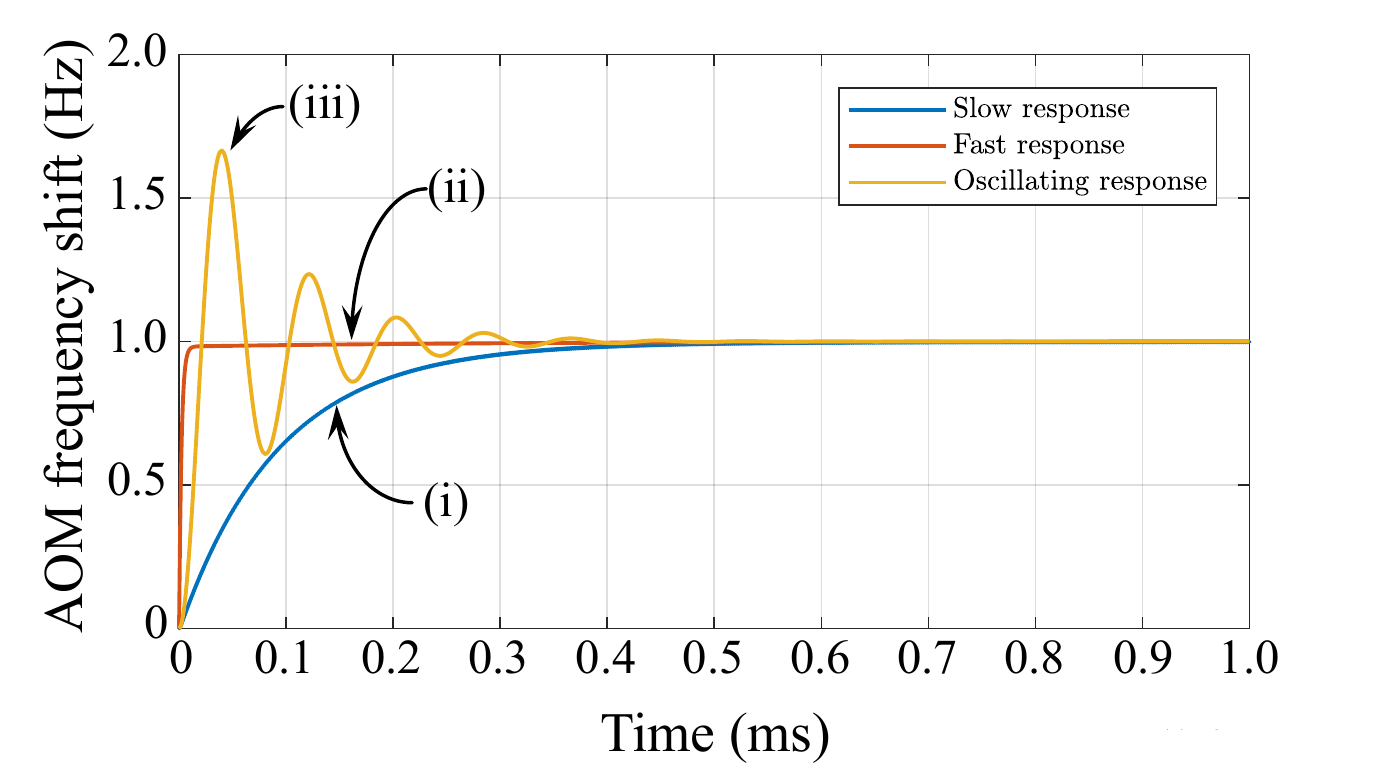}
\caption{\label{Fig_StepReponseSimu}
Simulation of the closed-loop response of the servo-loop to a step input. The three curves correspond to three different sets of gain coefficients: (i) $G_\mathrm{P}=1.2\times 10^6$, $G_\mathrm{I}=5.4\times 10^{9}\;$rad/s, $G_\mathrm{D}=57\;$s/rad, and $G_\mathrm{N}=1.2\times 10^6\;$rad/s; (ii): $G_\mathrm{P}=2.8\times 10^7$, $G_\mathrm{I} =7.2 \times 10^{10}\;$rad/s, $G_\mathrm{D}=2.4 \times 10^3\;$s/rad, and $G_\mathrm{N} = 5 \times 10^7\;$rad/s; (iii): $G_\mathrm{P} = 2.8\times 10^7$, $G_\mathrm{I} = 2.5 \times 10^{11}\;$rad/s, $G_\mathrm{D}=126\;$s/rad, and $G_\mathrm{N} = 7.7 \times 10^4\;$rad/s.}
\end{figure}

Figure \ref{Fig_StepReponseSimu} reproduces three different results for the simulation of the step response of the PDH servo loop for 3 different sets of gain coefficients of the PID controller, which are given in the figure caption. The other parameters of the simulation are : $G_\mathrm{PDH} = 1.9\;\mu$V/Hz, $f_{\mathrm{c,LPF}} = 1$~kHz and $f_{\mathrm{c,PD}} = 4$~MHz. The simulation calculates the time evolution of the frequency shift provided by the AOM when a Heavyside step is applied to the servo-loop input. One can see that, depending on the PID gain coefficients, the response of the servo-loop can be either (i) too slow, (ii) optimized, or (iii) oscillating.

\begin{figure}[h!]
\centering \includegraphics[width=0.8\textwidth]{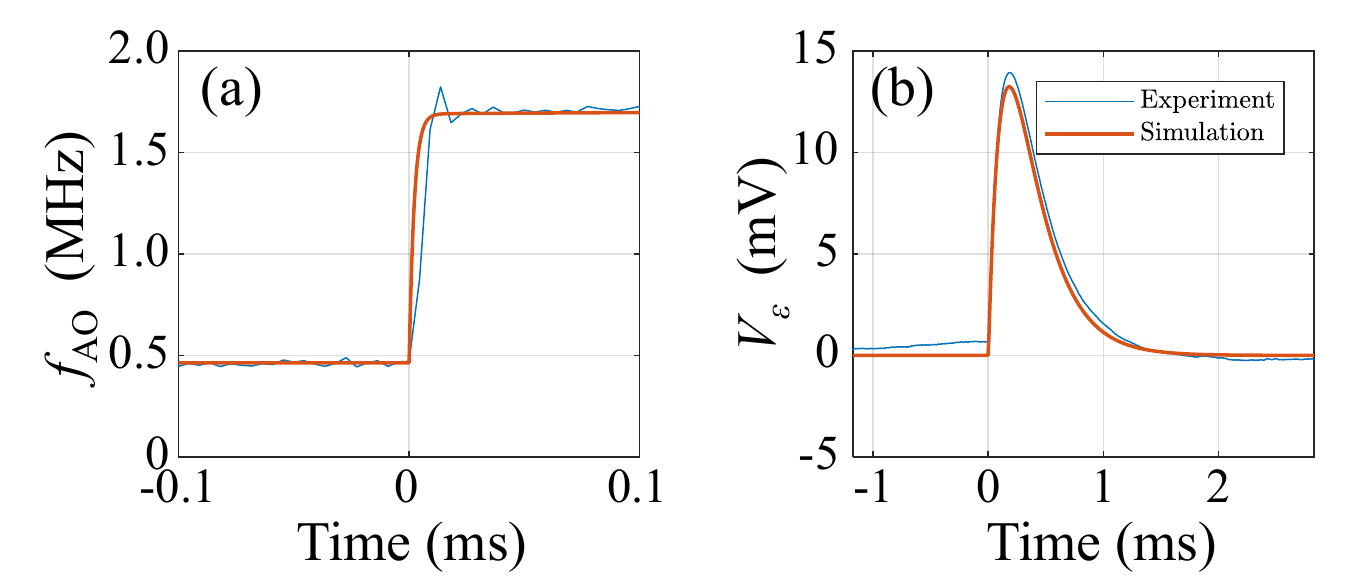} 
\caption{\label{Fig_SimuErretF} Time evolution of (a) the AOM$_1$ frequency and (b) the error signal $V_{\varepsilon}$ when a 500~kHz step is applied to the laser frequency. The PID gains are the ones corresponding to plot (ii) of Fig. \ref{Fig_StepReponseSimu}. The thin (resp. thick) lines correspond to the measurements (resp. simulations). The time origin corresponds to the moment where the frequency step is applied.}
\end{figure}

To check the validity of this model, we implemented such a PDH servo-loop based on the scheme of Fig.\;\ref{Fig_MOD_PDH} and used the optimized gain coefficients provided by the simulation, that should lead to a fast non oscillating response similar to the one of plot (ii) in Fig.\;\ref{Fig_StepReponseSimu}. 
To measure the response of the servo-loop to a frequency step, we introduce an extra AOM at the output of the laser, that we use to create a 500~kHz frequency step, while the servo-loop is closed. We then record the evolution of the AOM frequency and of the error signal with time. The results of this experiment are shown in Fig.\;\ref{Fig_SimuErretF}, and are compared with simulation results computed with the experimental parameters. The agreement between the experiment and the model is excellent, hence validating the modeling. Moreover, this shows that we are able to obtain a very stable frequency servo-loop, with an optimized closed-loop response time of the order of  10$\;\mu$s. Such a servo-loop is implemented for the two propagation directions in the next section.

\section{Rotation measurements and gyro performances}
\label{Rotation}
\begin{figure}[h!]
\centering\includegraphics[width=0.7\textwidth]{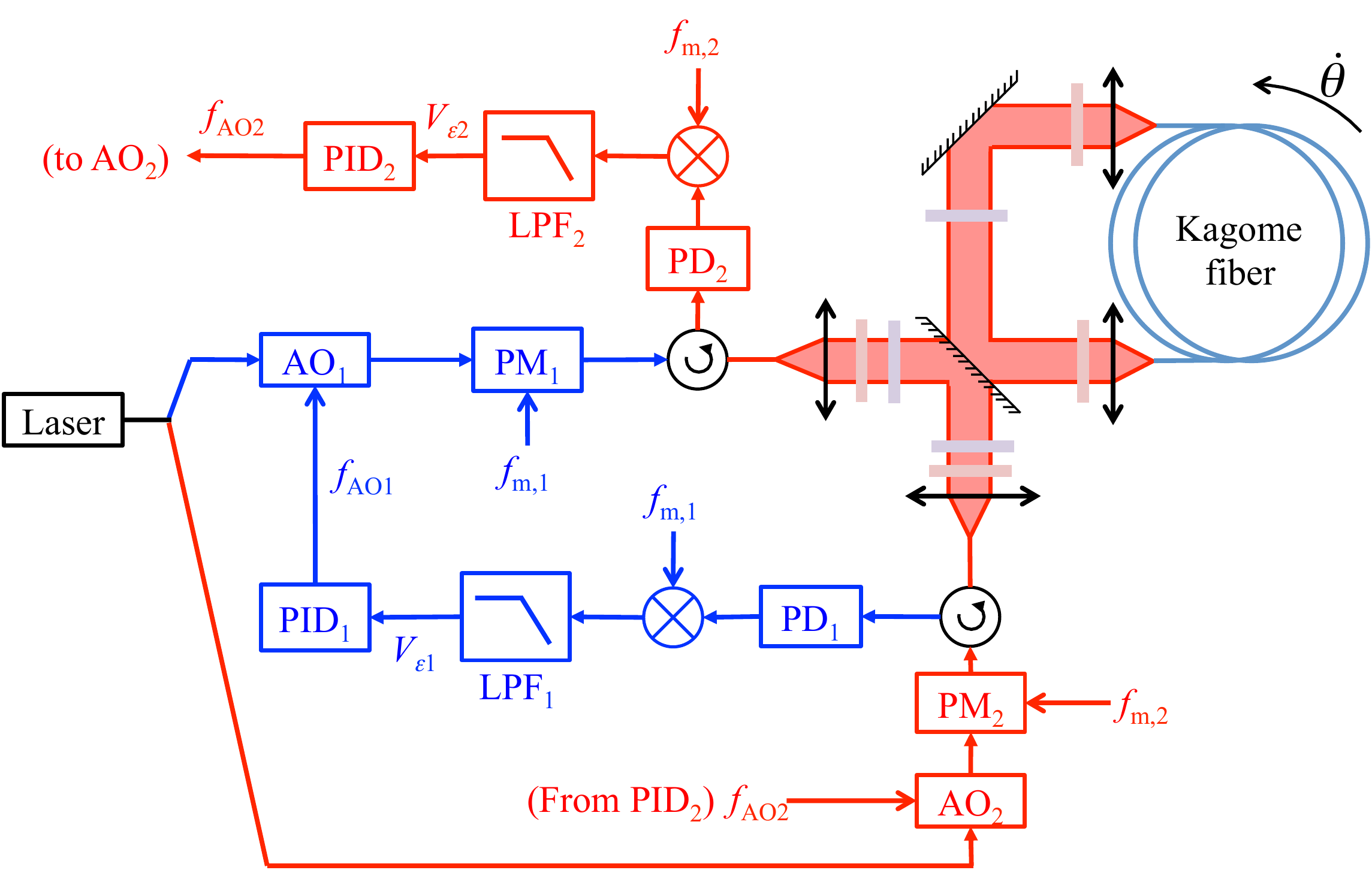}
\caption{\label{Fig_MontageMesureRot} Experimental setup aiming at measuring rotation rate. Two PDH servo-loops have been implemented to lock the laser to the two counterpropagating cavity resonances. The cavity is the one described in Sec. \ref{Sec_CaracCav}.}
\end{figure}

Figure \ref{Fig_MontageMesureRot} shows the configuration of our RFOG. The laser is split into two arms with a 50/50 fiber coupler. Two PDH servo-loops, labeled 1 and 2, similar to the one described in the preceding section, are implemented for the two propagation directions inside the cavity. The cavity is the one described in Sec. \ref{Sec_CaracCav}. The two propagation directions of the cavity are injected and the corresponding error signals are independently retrieved by means of two circulators. The two servo-loops permit to adjust the laser frequencies in the two directions thanks to two AOMs. In order to be able to independently retrieve the two error signals, corresponding to the two propagation directions, we choose two different phase modulation frequencies $f_\mathrm{m,1} = 3.012$~MHz and $f_\mathrm{m,2} = 2.89$~MHz.

Two important details are not visible in Fig.\;\ref{Fig_MontageMesureRot}. The first one is that the two AOMs use different diffraction orders. The first one has a central frequency of 110 MHz and diffracts light in  order $+1$, while the other one also has a central frequency of 110 MHz but diffracts light in order $-1$. This permits to excite two different contra-propagating longitudinal modes of the resonator separated by approximately 220~MHz, i. e. 13 times the cavity FSR, in order to avoid the lock-in phenomenon induced by  backscattering inside the cavity \cite{Aronowitz1999,Zarinetchi1986}. 

The second important experimental feature that is not described in Fig.\;\ref{Fig_MontageMesureRot} is the fact that we implemented two extra servo-loops in order to minimize the residual amplitude modulations created by the phase modulators PM$_1$ and PM$_2$. To this aim, in each propagation direction, 10\% of the light intensity that is about to be injected inside the resonator is sent to a photodiode. The detected signal is then demodulated at the corresponding modulation frequency $f_\mathrm{m,i}$ with $i=1,2$, filtered, and injected into a servo-loop that adds an extra voltage to the signal that feeds the phase modulator, thus permitting to minimize the residual amplitude modulation (RAM) \cite{ZhangW2014}. This minimization of the residual amplitude modulations of the two injected beam permits to strongly reduce the gyro bias drift.

\begin{figure}[h!]
\centering\includegraphics[width=0.8\textwidth]{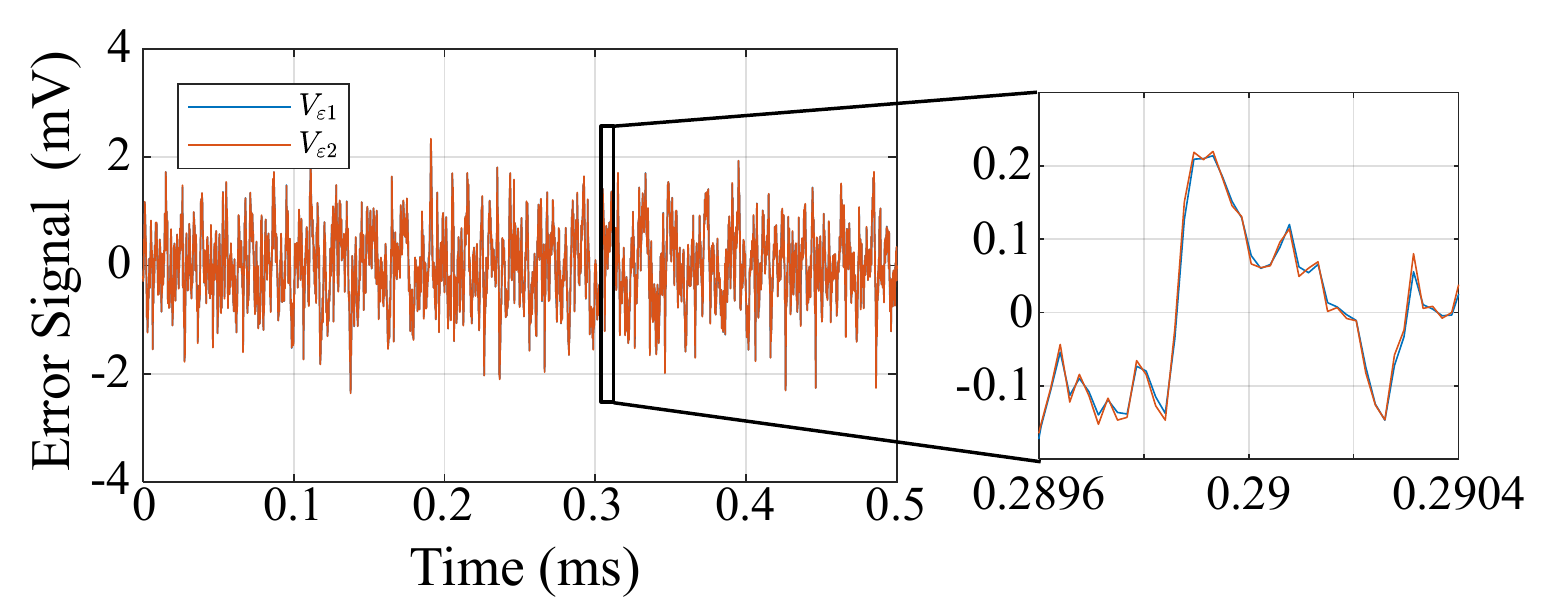}
\caption{\label{Fig_SignauxErreur} Time evolution of the two error signals when the two PDH servo-loops are closed. The right plot is a zoom on a section of the left plot.}
\end{figure}

Figure \ref{Fig_SignauxErreur} displays the time evolution of the error signals $V_{\varepsilon1}$ and $V_{\varepsilon2}$ when the two PDH servo-loops are closed, in the absence of any external rotation. As one can see, the error signals are strongly correlated. We indeed made sure that the value of $G_\mathrm{PDH}$ is the same for the two propagation directions by controlling the intensities in the two directions thanks to the AOMs. Having the same PDH gains, the same photodiodes, and the same demodulation filter cut-off frequencies allows to have the same PID gains for the two directions.

Once the servo-loops are closed and operating, the relation between the laser frequency, the frequencies of the AOMs, and the cavity FSR is:

\begin{equation}
\label{Eq_SagnacAO}
f_\mathrm{AO 1} + f_\mathrm{AO 2} = 13\;\Delta\nu_\mathrm{FSR} + \Delta\nu_\mathrm{S},
\end{equation}
where the Sagnac frequency difference is related to the angular velocity $\dot{\theta}$ through:
\begin{equation}
\Delta\nu_\mathrm{S} = \dfrac{4A}{\lambda nL}\dot{\theta} \equiv K\dot{\theta}\;.
\end{equation}
$A$ is the area of the cavity and $K$ the gyro scale factor.

\begin{figure}[h!]
\centering \includegraphics[width=0.9\textwidth]{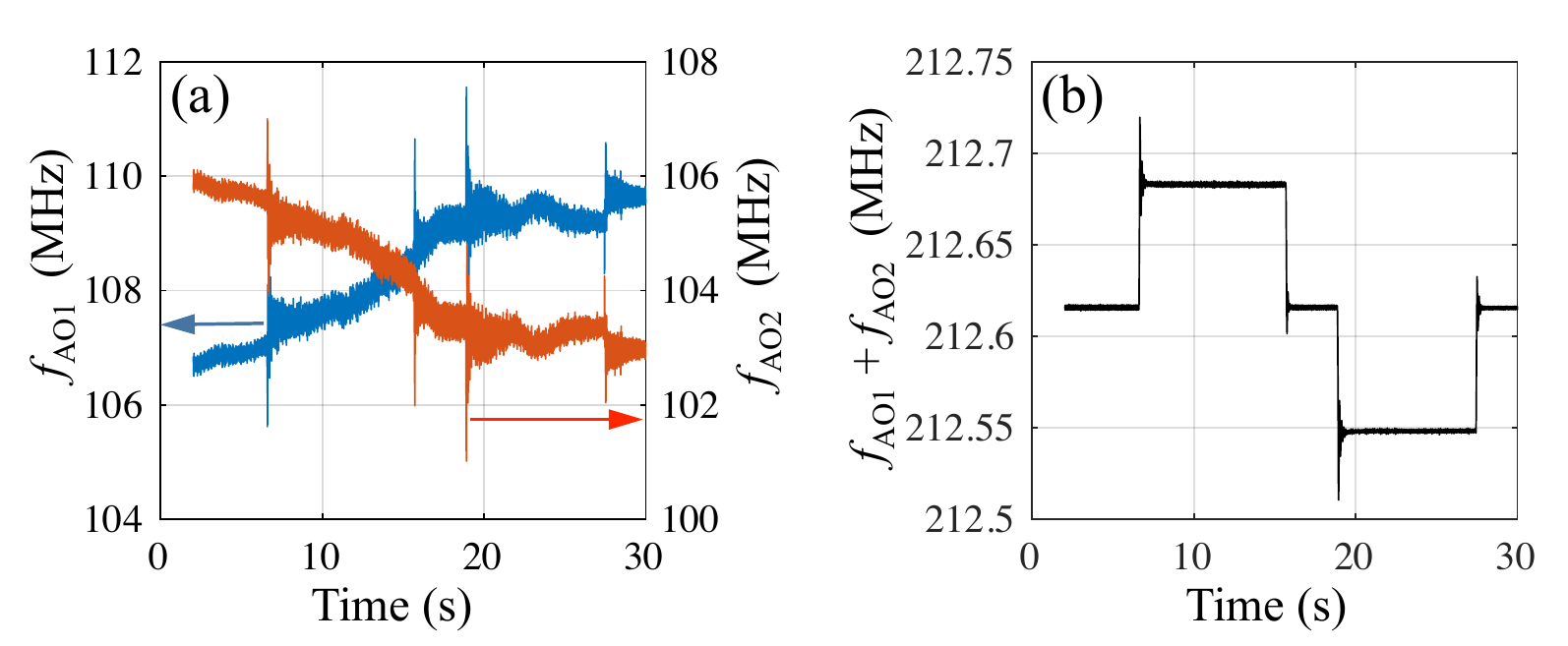} 
\caption{\label{Fig_FreqsRot10deg} (a) Time evolution of the frequencies of the two AOMs when the gyro is successively rotated at $\dot{\theta}=+10\,^\circ$/s and $\dot{\theta}=-10\,^\circ$/s. (b) Time evolution of the sum of these two frequencies.}
\end{figure}
We first performed a calibration of our gyro by applying different rotation rates and measuring the frequencies of the AOMs. To this aim, the sensing resonator is mounted on a controlled rotating table. Figure \ref{Fig_FreqsRot10deg}(a) shows the measured evolution of the frequencies of the two AOMs when the table is first at rest, then subjected to a $\dot{\theta}=+10\,^\circ$/s angular velocity, then at rest again, and finally rotating at  $\dot{\theta}=-10\,^\circ$/s angular velocity. As one can see, the AOMs frequencies drift with time, because of the drifts of the laser frequency and of the cavity length. However, when the frequencies of the two AOMs are added, according to eq.\;(\ref{Eq_SagnacAO}), one obtains the signal evolution shown in  Fig.\;\ref{Fig_FreqsRot10deg}(b). One can see that the drifts of two initial signals cancel and the rotation rate can be easily measured. As expected from eq. (\ref{Eq_SagnacAO}), the sum of the frequencies of the two AOMs is centered at $13\,\Delta\nu_\mathrm{FSR}$, which is equal to 212.61~MHz.

\begin{figure}[h!]
\centering \includegraphics[width=0.4\textwidth]{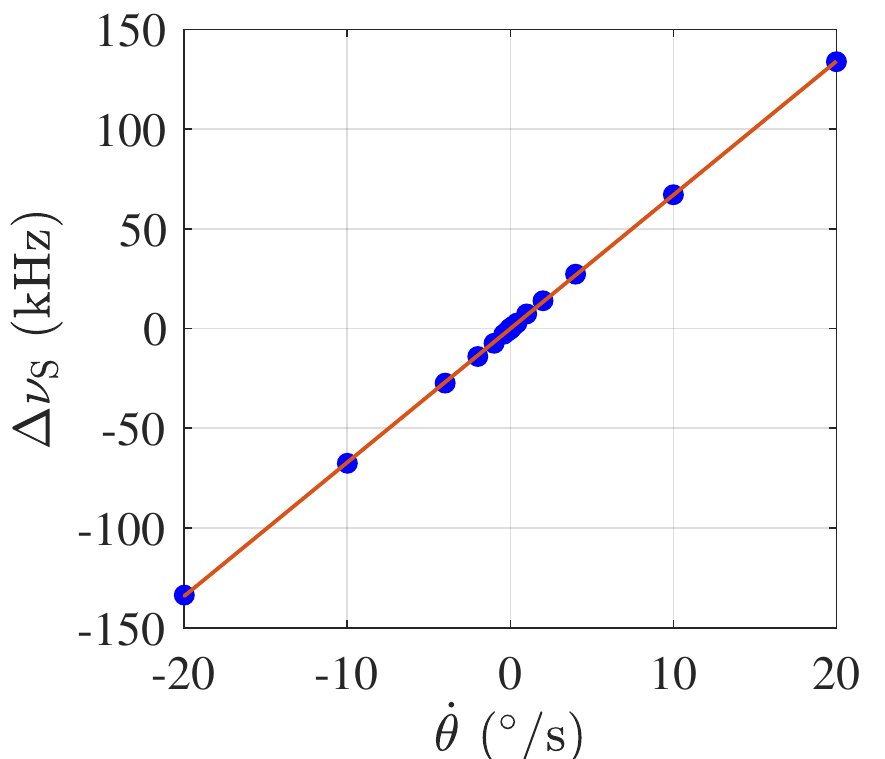} 
\caption{\label{Fig_Etalonagepm1} Calibration of the hollow core fibre based RFOG. Dots: measurements. Full line: linear fit.}
\end{figure}
By performing repeated measurements like the one of Fig.\;\ref{Fig_FreqsRot10deg} for various values of $\dot{\theta}$, we obtain the response curve of the RFOG reproduced in Fig.\;\ref{Fig_Etalonagepm1}. The response curve of the gyro is linear, as evidenced by the good agreement with a linear fit (full line in Fig.\; \ref{Fig_Etalonagepm1}). From this linear fit, we obtained a measured scale factor equal to $K = 6.67\;\mathrm{kHz/(^\circ/s)}$, in good agreement with the expected value $K = \dfrac{D}{\lambda} = 6.75\;\mathrm{kHz/(^\circ/s)}$ obtained for a diameter of $D = 60$~cm and a wavelength  $\lambda = 1.55\;\mu$m.

\begin{figure}[ht]
      \centering \includegraphics[width=0.95\textwidth]{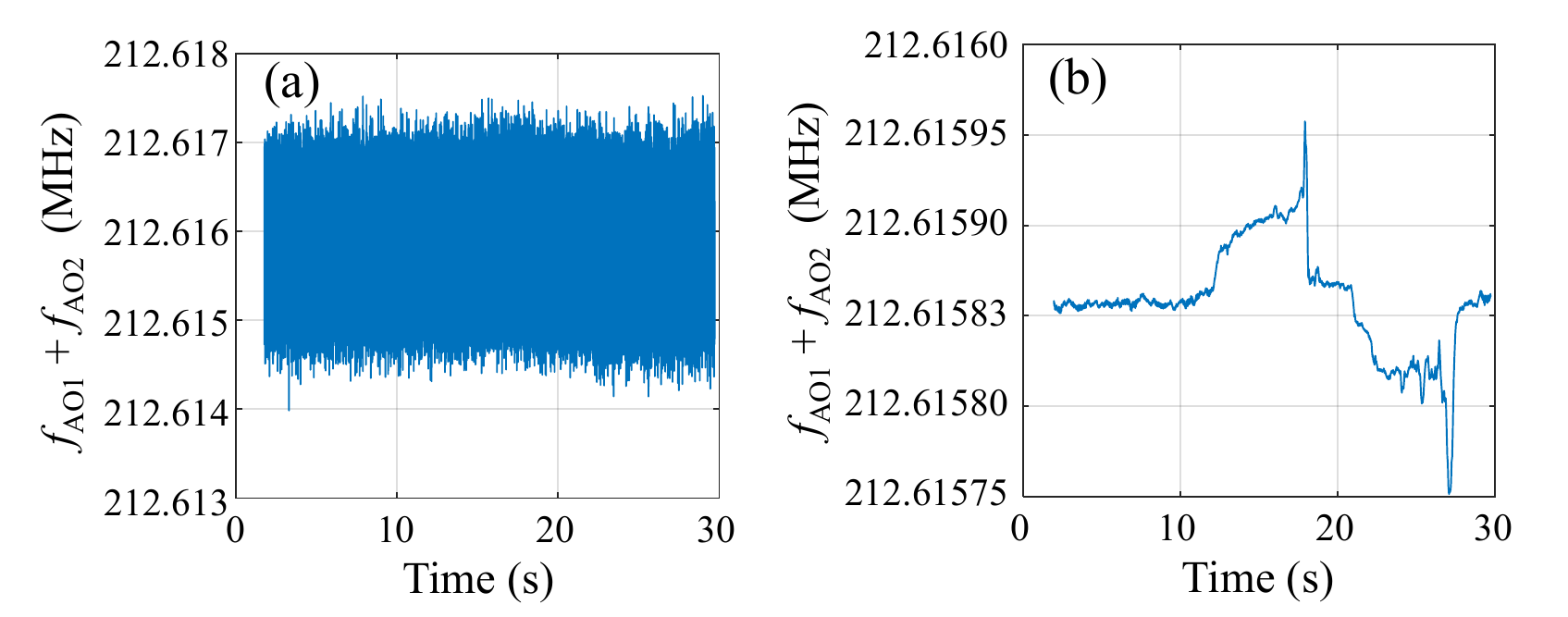} 
      \caption{\label{Fig_RotMinSignal} (a) Time evolution of the sum $f_\mathrm{AO 1} + f_\mathrm{AO 2}$ of the frequencies of the two AOMs. The rotating table carrying the gyro is rotated at  $\dot{\theta}=+0.005\,^\circ$/s, then stopped, and finally rotated at $\dot{\theta}=-0.005\,^\circ$/s. The Sagnac frequency shift is hidden by the noise. (b) Result of the digital filtering of the same signal with a 10 Hz bandwidth. The Sagnac effect is retrieved.}
\end{figure}

In order to probe the sensitivity of our RFOG, we monitored the AOM frequencies when the angular velocity of the rotating table is as small as $\pm\, 0.005 ^\circ$/s.  Figure \ref{Fig_RotMinSignal}(a) shows the evolution of the sum $f_{\mathrm{AO1}}+f_{\mathrm{AO2}}$ of the two AOM frequencies when the rotation rate $\dot{\theta}$ evolves from 0 to $+ 0.005 ^\circ$/s, then again to 0, to $- 0.005 ^\circ$/s, and finally back to 0. As one can see, the rotation rate is not detectable and is hidden in the noise. In the conditions of Fig.\;\ref{Fig_RotMinSignal}(a), the bandwidth of the gyro is estimated to be equal to 10\;kHz. We thus try to filter out the noise by reducing this bandwidth. Figure  \ref{Fig_RotMinSignal}(b) shows the result one obtains when the signal of Fig.\;\ref{Fig_RotMinSignal}(a) is digitally filtered with a 10-Hz-bandwidth filter. The $\pm 0.005 ^\circ$/s rotation rate is now clearly visible, with a signal-to-noise ratio of the order of 10. However, during the time interval in which the rotation rate is constant, clear drifts can be seen in the frequency $f_{\mathrm{AO1}}+f_{\mathrm{AO2}}$, showing that we are reaching rotations rates of the same order of magnitude as the bias drift of our gyro.

\begin{figure}[h!]
      \centering \includegraphics[width=0.8\textwidth]{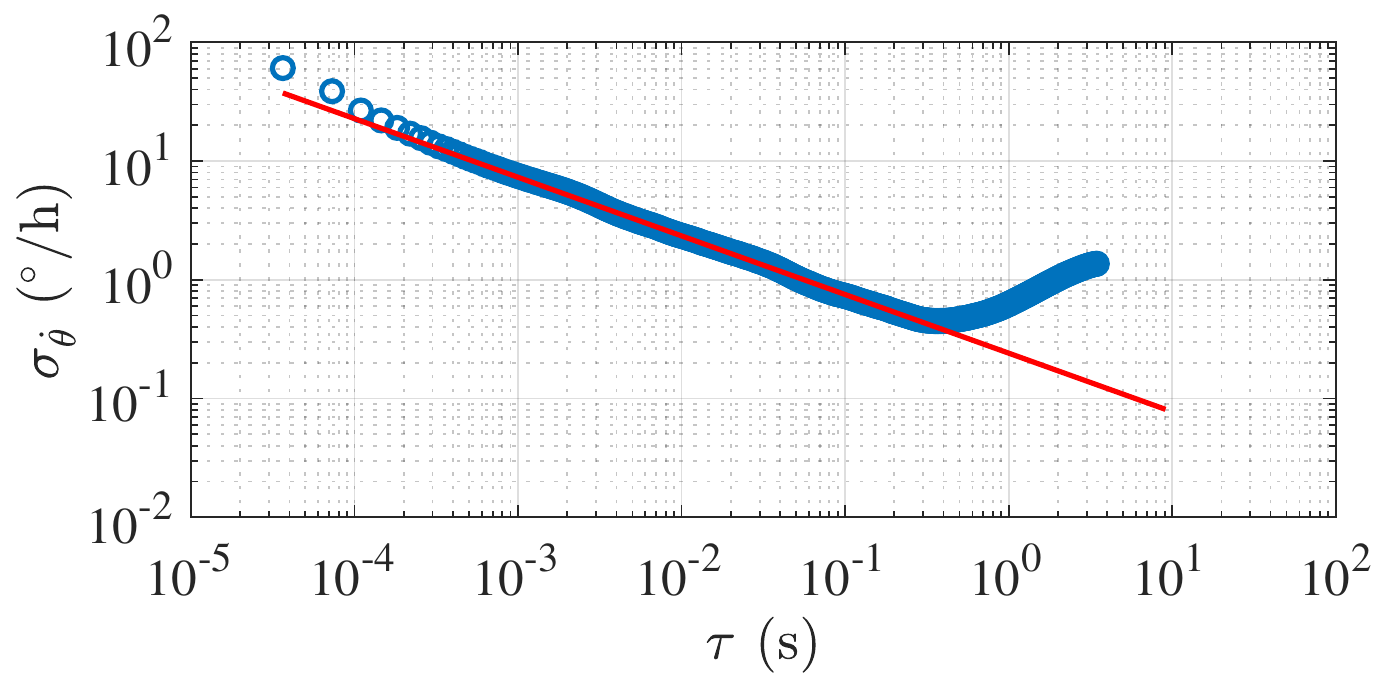}
      \caption{\label{Fig_AllanVar} Logarithmic plot of the evolution of the square root of the Allan variance $\sigma_{\dot{\theta}}$ of the measured rotation rate as a function of the integration time $\tau$. The linear fit of the first part of the plot has been obtained thanks to the following equation: $\ln(\sigma_{\dot{\theta}}) = -0.49 \, \ln(\tau) -1.42$.}
\end{figure}
Since the bias drift limits the values of the slowest rotation rates measurable by the RFOG, we calculate the overlapping Allan variance $\sigma_{\dot{\theta}}^2$ of the RFOG signal when it is not rotating \cite{Allan1966,Riley2008}. From the measurement of the evolution of this Allan variance as a function of the integration time, one can in principle deduce two characteristics of the RFOG : its Angular Random Walk (ARW) and its bias stability.

Figure \ref{Fig_AllanVar} shows the evolution of the square root of the  Allan variance $\sigma_{\dot{\theta}}$ as a function of the integration time $\tau$. This plot has been obtained thanks to an acquisition duration equal to 20\;s and a sampling period of $36\;\mu\mathrm{s}$. One can see that $\sigma_{\dot{\theta}}$ decreases with $\tau$ till $\tau$ reaches approximately 0.5\;s, after which $\sigma_{\dot{\theta}}$ starts to increase with $\tau$. The first part of the plot is fitted by a linear fit in the log-log representation, as shown by the red curve in Fig.\;\ref{Fig_AllanVar}. This fitted line has a slope equal to $-0.49$, which is in agreement with the $-1/2$ slope expected for an ARW induced by a white shot noise. It leads to a value of the ARW of our system equal to $\mathrm{ARW}=0.24$~($^\circ$/h)/$\sqrt{\mathrm{Hz}}$, which is equivalent to  $\mathrm{ARW}$=0.004~$^\circ/\sqrt{\mathrm{h}}$. This value compares very well with preceding results obtained using hollow-core fibers \cite{Terrel2012,JiaoHongchen2017}, showing that such a technology is a good candidate for navigation grade gyroscopes.  However, the ARW that we obtained is two orders of magnitude larger than the one expected from eq.\;(\ref{Eq_SNL}). This shows that our results are probably limited by other sources of noise than the shot noise, such as electronics noise. This makes room for further improvements in the future, based on more sophisticated electronics. Nevertheless, to the best of our knowledge, this is the best ARW result for a hollow-core fiber based RFOG \cite{JiaoHongchen2017}.

The minimum value of $\sigma_{\dot{\theta}}$ observed in Fig.\,\ref{Fig_AllanVar} corresponds to a bias drift equal to 0.45$^\circ$/h for a 0.5 sec integration time. This relatively large value of the bias drift must be attributed to the mechanical drifts of our cavity setup, which is not well isolated from mechanical and acoustic perturbations.

\section{Conclusion}
 In conclusion, a RFOG based on a hollow-core Kagome fiber resonator was built and tested. It is based on a 18-meter-long fiber in a semi-bulk architecture using mirrors to close the 60-cm-diameter resonant cavity. The gyro was operated in closed-loop regime with two PDH servo-loops monitoring the variations of the cavity resonance frequencies in the two counterpropagating directions.  Moreover, the two counterpropagating waves were locked to two different longitudinal modes in order to avoid any lock-in at low rotation rates, and the residual amplitude modulations of the phase modulators were actively controlled. We then obtained an ARW equal to $0.004^\circ/\sqrt{\mathrm{h}}$ and a bias stability equal to $0.45^\circ$/h.
 
 The results reported here open the way towards the realization of a hollow-core fiber based RFOG with navigation grade performances, provided the mechanical stability of the resonator can be improved to decrease the long-term bias drift, and provided one can find a fiber that can tolerate a smaller radius of curvature to decrease the size of the resonator. For example, the recent development of negative-curvature fibers \cite{Gao2018} could meet these requirements.

\section*{Acknowledgments}
This work is supported by the Agence Nationale de la Recherche (Project PHOBAG: ANR-13-BS03-0007), the European Space Agency (ESA), and the R\'egion Nouvelle Aquitaine. 
The work of AR, GF, and FB has been performed in the framework of the joint research laboratory between Laboratoire Aim\'e Cotton and Thales Research \& Technology. The authors are happy to thank Matthieu Dupont-Nivet for his help with the Allan variance.
\bibliography{Biblio}






\end{document}